\documentstyle[11pt]{article}
 
\def\pa{\partial}
\def\k{\kappa} 
\def\g{\gamma} \def\G{\Gamma}
\def\a{\alpha} 
 
\def\d{\delta} \def\D{\Delta}
\def\e{\epsilon} 
 
\def\k{\kappa}
 \def\L{\Lambda}
\def\m{\mu} 
\def\n{\nu}

\def\mn{{\mu\nu}}
     
\begin{document}

\begin{flushright}
BRX TH-435
\end{flushright}

\begin{center}
{\large\bf D=11 Supergravity Revisited}

S. Deser\\

Department of Physics\\
Brandeis University\\
Waltham, MA 02254, USA
\end{center}

\begin{quotation}
I discuss two novel results in D=11 Supergravity.
The first establishes, in two complementary ways,
a no-go theorem that, in contrast to all D$<$11,
a cosmological extension of the theory does not
exist.  The second deals with the structure of
(on-shell) four-point invariants. These are
important both for
establishing existence of the lowest (2-loop) order
candidate counter-terms in the
theory proper, as well as for comparison with
the form of eventual
``zero-slope"  QFT limit of M-theory.
\end{quotation}

\noindent{\bf I. ~Introduction}

It is a particular pleasure for me to be present on
this occasion to celebrate Dick Arnowitt's 
?-birthday.  Although not (quite) Dick's
oldest collaborator in age, I have seniority in
terms of years: our first publication was in 1953,
45 years ago, and our joint work began a couple
of years before that, with an (as yet) unpublished 
manuscript.
In the twelve year span between 1953 and
1965 we wrote some
30-odd papers, and (about) 85\% of a book on
general relativity, which I find useful in teaching
to this day!  I am also happy to see 
other (mutual) old collaborators here,
including (in time ordered sequence)
Charlie Misner, Mike Duff and  Bruno Zumino,
as well as other TAMU friends, to whose work
I will in fact be referring.

There have been many changes in relativity since
our old days; for one thing, the size of expert
audiences has greatly increased as that subject
moved towards center stage.  For another, ``ADM",
originally regarded as a disreputable intrusion
of
quantum field theoretical ideas into classical gravity 
has ended up  as an
acronym whose meaning is barely remembered (in either
camp), 
a sign of true acceptance!  Since those days, I
(unlike Dick) have strayed from the real world, and
find myself currently in D=11, (supergravity, to boot),
about which I will
be speaking here.  Supersymmetry itself has of
course motivated much of Dick's work from its
earliest days; he just persists in believing
 we live in D=4.

\noindent{\bf II.~ D=11 Supergravity: Uniqueness}

Supersymmetry, both as a global invariance but 
especially in its local, supergravity, context, is
perhaps the most powerful and ubiquitous single
invariance principle to have emerged in the past
twenty years; it seems to underlie a wide variety
of seemingly different phenomena, including, most
recently, the dualities that have unified hitherto
separate superstring models into a single M-theory.
Sometimes, the very ``threat" of supersymmetry is
sufficient to reestablish deep results such as
positivity of gravitational energy.  While 
the mathematical tools that physicists use in
supersymmetry are neither new nor complicated --
basically Grassmann variables and Clifford
algebras -- yet there is clearly a lot left to
understand in the unreasonable success of 
supersymmetry in physics. 
I believe that we still do not fully grasp 
at an intuitive level why the
existence of a ``Dirac square root" brings so many
amazing ``coincidences", cancellations of everything
from ghosts to infinities, uncanny dualities
and even a preferred
spacetime dimensionality.  This is not the place to
go through the vast literature of any one of the 
subsections of supersymmetry.  Instead, I will 
confine myself here to some novel aspects of
what is in some ways the quintessential super-system,
D=11 supergravity (``Sugra").  I remind you of some basic
history: Sugra, first discovered 
in D=4, followed by ``degenerate" versions in D=2 
(superstring), D=1 (superparticle), and
D=3 (supermembrane), rapidly made its way up the 
dimensional ladder \cite{001}, 
ending with the ``ultimate"
rung of D=11 \cite{002}.  The reasons for this
ceiling were in fact mathematically prosaic ones
having to do with properties of Clifford algebras
\cite{003,001} in signature (D--1,1),
but also reflected physical
requirements that no massless fields with spin
greater than 2, and no more than one graviton, be
permitted.  The former is due to the
incompatibility of gravitational interactions of
higher spin gauge fields \cite{004} which has long
been ``understood"; more than one graviton 
is more intuitively seen to be a bad thing,
but that can also be formalized.

Unlike its lower dimensional manifestations, D=11
Sugra was also seen to be our ``uniquely" unique
QFT, in the sense that its hallmark
requirement -- equality of bose and fermi modes
-- necessarily adjoins to the graviton a single
spin 3/2 fermion together with a (singlet) 3-form 
gauge potential $A_{\mu\nu\alpha}$, with neither 
``N$>$1" extensions nor matter coupling permitted.  
Apart from the usual proliferation of 4-fermion 
terms (and one nonminimal coupling term), 
the action is simplicity itself, schematically 
\begin{equation}
I = \int d^{11} x
[\kappa^{-2} R + \bar{\psi}_\mu  
\Gamma^{\mu\nu\alpha} D_\nu \psi_\alpha +
F^2_{\mu\nu\alpha\rho} + \kappa A \wedge F \wedge F]
\end{equation}
where $F$ is the four-form curl of $A$, $D_\nu$ is the
covariant derivative, and $\kappa^2$ is the
Einstein constant with obvious dimensions $L^9$ in
$\hbar = c = 1$ units.  Amusingly, the 
(metric-independent) Chern--Simons
term in (1) seems to have been its first 
physics appearance, followed by similar ones on
all lower (odd) $D$; it is
(uncharacteristically) $P$ and $T$ even.
It was not for another few years that such terms would
begin to emerge in the more familiar QED$_3$
context.

I remind you that the degree of freedom count
for Einstein gravity is D(D--3)/2=44, the
number of transverse-traceless spatial metric 
components, that of the form field counts
the transverse spatial components $A_{ijk}$
(invariance is under $\d \, A_{\m\n\a} =
\pa_{[\m} \xi_{\n\a]}$), so it is
 (D--2)(D--3)(D--4)/3! = 84, while the fermionic
spinor-vector has (D--3) transverse and 
$\g$--transverse for the spatial vector index
times the usual Majorana spinor count
$2^{[D/2]-1} = 128$ (the $\frac{1}{2}$
is for first-order).  The corresponding 
invariances are, symbolically
\begin{eqnarray}
\d\, e_{\m a} & = & \bar{\a}(x) \g_a\psi_\m
(x) \; , \hspace{.4in} 
\d\psi_\m \; = \; [D_\m + (\G F)_\m ]
\a (x) \nonumber \\
\d\, A_{\m\n\a} & = & \bar{\a}(x)
(\G\psi (x) )_{\m\n\a}
\end{eqnarray}
where $\G^{\m{_1} \ldots \m{_n}}$ is a suitable
$n$-index ``gamma" matrix, $e_{\m a}$ the vielbein 
and $\a (x)$ a Grassmanian parameter.

Uniqueness up to now has meant that, 
given the Einstein action as the geometrical term of 
the system, the rest of (1) necessarily 
follows.  For example, even replacing the
4-form $F$ by its equivalent dual 7-form does not
lead to a consistent formulation despite its
seemingly respecting the degree of freedom count.  
Indeed, there is not even any D=11 globally
SUSY matter (highest spin $<$2)
system, hence no sources of 
(1). As I said, this is in sharp contrast with all
lower dimensional cases, including the ``nearest
neighbor" D=10.  What I want to talk about 
here is recent work \cite{005} on the remaining
possible non-uniqueness, involving the replacement 
of the local
Lorentz group underlying the Einstein gravity of 
(1) by the (anti) de Sitter one, through the 
introduction of a cosmological term
$\Lambda \int d^{11}x \sqrt{-g}$ for
gravity -- Einstein's ``biggest" but unavoidable,
``mistake".  [As we know all too well, such a term
arises always in ordinary QFT coupled to
gravity, with a horribly wrong natural magnitude;
originally, there was hope from the fact that this
zero point energy is absent in supersymmetric
QFT's, but only if supersymmetry is unbroken.]  
In any case,
the rapid construction of supergravities with
cosmological terms (of anti-de Sitter sign)
\cite{001} undercut this hope; indeed such models
were possible for all dimensions in addition to the
original D=4, including D=10.  However, the 
apparent exception was D=11: on the one hand,
arguments based on Clifford algebra
seemed to forbid any simple such extension, but the
more general ``no-go" question remained open for a
long time.  Once D=11 supergravity reclaimed its 
rightful place, as the QFT limit of M-theory, and was 
no longer the enigma in a world of D=10 superstrings,
it became more important to settle it.  
I will now briefly sketch the two ways we used to do so, 
and refer to \cite{005} for details. Our negative upshot 
means that D=11 Sugra is, beyond all its other amazing
properties, the only QFT we know that forbids the
presence of a cosmological term, and does so
because of supersymmetry.  Breaking the latter
by simply including this term is forbidden by
consistency considerations, unless of course one
admits truly massive $\psi_\m$ and $A$ fields.

Cosmological Sugra, when it does exist (for D$<$11),
is based on a simultaneous extension of gravity and
gravitino actions, the former with the usual
$\int \L \sqrt{-g}$ term, the latter with a 
``mass" term where $m \sim \sqrt{-\L}$ is what
requires adS sign for $\L$.  The reason this 
simultaneous deformation (at least)  is needed
is that since small
gravitational excitations $h_{\m\n}$ about the
vacuum (here adS) have the same excitation count as for
$\L$=0, so must the fermions. The quadratic 
graviton action is still gauge
invariant under $\d h_{\m\n} = \bar{D}_\m 
\xi_\n + \bar{D}_\n \xi_\m$ where $\bar{D}_\m$ 
is the covariant derivative with respect to the 
background.  This is
precisely what is made possible by the mass term:
gravitino excitations about adS also maintain their
usual flat space gauge invariance, but with
$\d \, \psi_\m = (\bar{D}_\m + m \, \g_\m ) \e
(x) \equiv {\cal D}_\m \e (x)$, because (only)
these extended covariant derivative commute,
$[{\cal D}_\m , {\cal D}_\n ] = 0 $ if  the 
mass is also ``tuned" to adS as noted above.

1.~ The Noether way.  One of the most useful, if
seemingly pedestrian, tools we have had in building
up nonabelian gauge theories from abelian ones -- 
and also seeing when that is not possible -- is the 
Noether procedure.  Here one tries to gauge the 
simple linear theory by self-coupling its 
conserved current (if any!) in a possible infinite series
of steps to reach a consistent nonlinear one.  This 
is how one can get from Maxwell to Yang--Mills or
from spin 2 fields to Einstein gravity or from
spin 2 plus spin 3/2 to supergravity \cite{006}, 
but not, for example from spin 2 plus 5/2 to any 
consistent interacting model. To be sure, the
starting point must be commensurate with the
desired end: one cannot reach 
general relativity with a cosmological constant from
the free theory in background flat space, but only
if one starts with the free theory in a de Sitter 
(or adS) background.  Here then, the challenge
is to start with the assembly of  free bose and fermi
constituents in the AdS context, look for a
Noether current associated with their global
supersymmetry and attempt to bootstrap to the
desired local invariance.  That is, we try to
mimic the way the known correct local Lorentz
supergravity can be reached from its corresponding 
non-interacting components.

We already know how to start the linearized gravity and
gravitino systems off as free gauge systems in the
background that keep the correct bose-fermi 
equality; what 
about the form field?
Because it is a form, it only depends on curls
which of course do not change in nonflat geometries,
so its count is already safe.  Indeed,
the only possible deformation here would also be a
mass term $\sim m^2 A^2$, but (unlike its gravitino
counterpart) that would break 
the invariance and therefore unacceptably raise the 
form field's excitations from 84 to 120.  
So we have the desired starting point, three
linearized systems so defined that their excitation
content is correct also in the background.  Can we
define an initial ``global SUSY" transformation for
them?  This is {\it a priori}
possible, because there is a ``constant", Killing,
spinor $\a (x) $ such that 
${\cal D}_\m \a = 0$, consistent
with $[{\cal D}_\m , {\cal D}_\n ] = 0$.  
I emphasize that this ``constant" $\a$ is 
{\it not} the same as the nonconstant 
$\e (x) ({\cal D}_\m \e \neq 0)$
under which the pure fermionic
system is invariant by itself!
So there is a candidate tranformation, 
but it is not an
invariance because of the $F$-field.  What
happens is that the $m \bar{\psi}\psi$ term we had
to introduce to maintain the ``internal" gauge
invariance of the gravitino action necessarily
varies, though the global supersymmetry parameter
$\a$ into a term that cannot, already on
dimensional grounds, be cancelled by varying
$F^2$, nor can we usefully alter its natural
$\d A \sim \bar{\a} \G \psi$ variation.
So the form field is the obstruction to so much
as even an initial Noether current and there is 
no ``zeroth" step.

2. ~Cohomology. This approach is complementary
to the first; it is better suited to 
a different  starting-point,
the full $\Lambda = 0$ Sugra of (1).  Suppose
we immediately accept the full $\L =0$ action
(1) as the starting-point of the desired extension,
and look for a consistent deformation of this full
nonlinear model with its ``nonabelian" gauge 
invariance, that will include a cosmological term.
More precisely, since this term is necessarily
associated  to a mass term
$\sim  m\bar{\psi}_\m \G^{\m\n} \psi_\n$, 
$m \sim \sqrt{-\Lambda}$
for the fermion as explained earlier, we begin the
deformation process with terms linear in
$m$, the cosmological one acting as a second 
order deformation.  The beauty of the cohomological
description is that we need not separately adjust
the action and the transformation rules.  If the
deformation process is at all possible, it will
reveal itself at one go.  Here it is again the 
form field that blocks the process and forbids
any extension of the desired type. For all lower 
$D$, consistent deformations exist. [This
obstruction is also true with a dual 7-form 
description.]  However, here if we adjoin to the 
original system $S_0$ (including ghost completions)
a $\D S_1 \sim m \int \bar{\psi} \psi$, we find that 
we cannot even maintain the first order
consistency $[S_0 , \D S_1]=0$ let alone use
$\D S_2 \sim \L \int \sqrt{-g}$ to cancel
$[ \D S_1, \D S_1]$ with $[S_0 , \D S_2]$.
Thus both approaches tell us independently that 
there is no extension of D=11 supergravity that
contracts back to it.  

It should be emphasized that, like all no-go
theorems, ours is predicated
on some assumptions that we believe to be 
reasonable; in particular,
that  
~a) the $m\rightarrow 0$ limit must be smooth
(as for D$<$11), and ~ b)~ no new dynamics beyond
our three initial fields enters. 
When supergravity is broken
or we compactify down to lower dimensions,
$\L$ can of course reappear!

\noindent{\bf III. ~D=11 Supergravity: On-shell 
Invariants}

My second topic is the construction of on-shell
invariants in D=11, and is still a work in progress
\cite{007}.  The motivation is twofold: First,
to determine the possible local counterterms 
that can be constructed, {\it i.e.}, at what
loop order does the theory begin to (or at
least is able to) pay the price of 
depending on the (dimensional) Einstein coupling
constant?  Second, and potentially more 
important is to thereby discover what corrections
to this limiting corner of M-theory should be 
sought, much like determining corrections
predicted by string theories to their zero-slope 
QFT limits. Unfortunately this is hard 
work because no
formalism exists at D=11 to generate such 
invariants and a more arduous road, using
physical arguments is needed. I shall only 
sketch our approach in the following.

There is one guaranteed way to generate
an invariant in any theory: Consider the tree
level amplitude (so no regularization worries
appear) for some specific number of particle 
scattering, say the 4-point functions.  All
external legs are real, so we are 
``on-shell" for the invariants that express this
amplitude.  Thus at lowest order, at least, global 
supersymmetry is preserved by the effective
action that expresses these amplitudes.
This is of course a statement that the
program must succeed, but not yet a 
concrete result.  What enables us to
proceed, apart from an awful lot of 
calculation, is the ability to cast the 
primitive scattering graphs, such as
graviton-graviton or form-graviton, into
expressions that are written entirely in terms
of Riemann, or better, Weyl tensors for the
gravity part.  Here previous experience \cite{008}
in D=4 tells us that the Bel--Robinson (BR)
tensor \cite{009} will play an essential role,
which helps.  Another expected ingredient is
the famous expression \cite{010} of
(D=10) string theory zero slope corrections
to D=11 supergravity involving terms like
$t_8\:t_8 \;\; R_1\ldots R_4$, where $t_8$
is an 8-index quantity made out of Kronecker
deltas and the $R$'s represent Riemanns or Weyls
whose indices they contract.  The link between 
all those scalars can be obtained
by means of another TAMU work \cite{011}, the
exhaustive enumeration of quartic curvature 
invariants.

So the flow chart is more or less as follows:
take all 3- and 4-point vertices in (1).  Sticking to
the bosonic sector, all we need are the 
$\k h^3$ and $\k^2 h^4$ gravity terms
$(\k h_\mn \equiv g_\mn -
\eta_{\m\n}$ and I omit showing
derivatives), the 3-point $\k h_\mn T^\mn
(F,g=\eta )$ and 4-point
$\left. \k^2 hh \frac{\d T}{\d g} 
\right|_{g=\eta}$ mixed vertices and
finally the Chern--Simons 3-point $\epsilon AFF$
interaction itself.  Now draw all possible tree
diagrams using all these vertices, contracting the 
intermediate graviton or form propagator
to a point.  [Technically, this is all done
in a systematic way in terms of the 
Mandelstam variables $s,t,u$.]  The
contact vertices ($hhFF$ and $hhhh$) are
just there to keep gauge invariance 
(Ward identities) honest.  So basically we
have $(hh)(hh)$ factorization of the 4-graviton
amplitude, say, into two graviton ``stress tensors".
While the latter cannot be quite well-defined 
(you heard it first from ADM 
\cite{012})
the contact terms save the day -- as we know in
the end they must!\footnote{This is a sort of
realization of a notorious problem in MTW
\cite{013} relating the Bel-Robinson (BR) tensor
to the graviton stress tensor's double derivative.}
So we will be able, using the $(s,t,u)$ derivatives
that appear in the amplitude, to provide the
correct $R^4$ four-graviton effective action,
presented as the sum of squares of BR-like 
currents. [But note that in this
D$>$4 context, there is more than one of 
those!] At D=4 there is only one BR and the action
reduces precisely to
the famous maximally helicity-violating
combination $(E^2_4 + P^2_4)$ where
$(E_4, P_4)$ are the Euler and Pontryagin
densities \cite{014}.  It also agrees with the
$t_8t_8\;R^4$ term there, taking into account that
to this (lowest) quartic order in $h$, 
the D=8 Euler density
$E_8$ is a total divergence in {\it any} dimension,
not just D=8.  The matter (4-form) 
contributions can also be uniquely obtained for 
both the $F^2R^2$ and $F^4$ amplitudes
(there is also a ``bremsstrahlung" $F^3R$
possible contribution representing graviton
emission from some leg of the CS vertex).  
Strictly speaking, one should check global 
supersymmetry of the resulting expression, 
but that is of course guaranteed by our
construction, and would only serve as a check
on our arithmetic of the various coefficients.
The reason it is hard to do explicitly is that it
first requires knowledge of all amplitudes involving
two gravitinos and two of our bosons, a possible
but unattractive calculation.  The internal checks
on the pure $R^4$ terms as well as the BR
structures are really sufficient.

What is all this good for?  There are two --
equally important -- applications:

First: is D=11 supergravity perhaps a 
miraculously finite theory?  It can't be just
renormalizable with its dimensional coupling
constant $\k$, so it is either nonrenormalizable
as lower dimensional Sugras are or all its
candidate counterterms must vanish for some
unknown reason.  Now our construction shows
that already at lowest possible (two-)
loop order there 
is such an invariant.  Whether its coefficient 
vanishes upon explicit calculation is of course a
separate question, but apart from the expected
absence of random cancellations (just as for 
2-loop pure gravity in D=4 \cite{015}) there
is a remarkable new development.  In a very
recent paper, Bern, Dixon, and collaborators
\cite{016} have reduced supergravity loop
calculations to super Yang--Mills ones in a very
powerful way.  Extrapolating their work beyond the
D=10 barrier to super-matter systems 
then strongly suggests
that this term does appear as a counterterm
and hence dashes 
any hope that D=11 Sugra is different from
the corresponding
lower D ones, all of which seem to go bad 
\cite{016}.  Let me explain
parenthetically why 2 loops:  In our expansion
in powers of $\k^2$, tree level being
$\sim \k^{-2}$,  one-loop term would be
$\sim \k^0$.  But there is no possible local
counterterm $\D I_1 = \int d^{11}x \: \D L_1$, 
since this would require an odd number of 
derivatives \cite{017} (all this is of course
in terms of dimensional regularization).
The only one-loop candidate of 
dimension 11 is the Chern--Simons one, 
$\sim \e \G \: RRRRR$ which is parity-odd.
At 2 loops we have $\k^{+2}d^{11}x$, so
$\D L_2$ must have dimension 20, {\it e.g.},
$\D L_2 \sim R^{10}$ or 
fewer $R$'s and more derivatives, like
$ R^4 \Box^6$
where $\Box^6$ is symbolic for derivatives
acting on the curvatures.  [In D=4, the
3-loop term $\sim \k^4 \int d^4x \: R^4$
was the lowest possible one
\cite{008}, there being no 1- or 2-loop
invariants available.]

The second application is in a way still more
interesting, because it should find direct
application in testing corrections of M-theory
to its D=11 Sugra limit, somewhat like the
zero slope corrections of string theory we 
mentioned earlier gave $R^4$ additions
to D=10 Sugra (but not of course D=11
directly!)  That is, whatever the right
M-theory may be, it should not only reduce
to this field theory, but produce additional 
effects necessarily 
starting with the above invariant this time as a
{\it finite} correction.

Apart from the intrinsic value of these
applications, I 
should add that learning to deal even with the
purely gravitational sector has also taught us 
a number of
hidden properties of (tree-level) 
general relativity, such as how
its diffeomorphism invariance translates into 
the gauge-invariant structure of physical scattering  
amplitudes.  This is just the sort of question
that ADM were in fact aiming for
(those curvatures are just glorified ``$h^{TT}$'s")
long before supergravity came on the scene!

\noindent{\bf Acknowledgements}

It is a pleasure to
thank my collaborators K. Bautier, M. Henneaux
and D. Seminara.  This work was supported by
NSF grant PHY 93-15811.

\end{document}